\documentclass[preprint,12pt]{elsarticle}

\usepackage{amssymb}
\usepackage{amsmath}
 \usepackage{xcolor}

\journal{Journal of The American Ceramic Society}

\begin{document}
\begin{frontmatter}



\title{Oxyfluoride glasses obtained through incorporation of CaF$_2$ into photovoltaic cover glass melts} 


\author[UTFPR]{Rafaela Valcarenghi}
\author[UEM]{Brenno Silva Greatti}
\author[UEM]{Robson Ferrari Muniz}
\author[UEM]{Vitor Santaella Zanuto}
\author[UTFPR,IFPR]{Anna Paulla Simon}
\author[Toledo]{Ricardo Schneider}
\author[UTFPR]{Raquel Dosciatti Bini} 
\author[UTFPR,UNO]{Márcio Antônio Fiori}
\author[CEMHTI]{Maxence Vigier}
\author[CEMHTI]{Emmanuel Veron}
\author[CEMHTI]{Mathieu Allix}
\author[UTFPR]{Marcelo Sandrini} 
\author[UTFPR]{Marcos Paulo Belançon}

\affiliation[UTFPR]{organization={Universidade Tecnológica Federal do Paraná},
            addressline={Via do Conhecimento Km 01}, 
            city={Pato Branco},
            postcode={85503-390}, 
            state={Paraná},
            country={Brasil}}

\affiliation[UEM]{organization={Universidade Estadual de Maringá},
            city={Maringá},
            state={Paraná},
            country={Brasil}}

\affiliation[Toledo]{organization={Universidade Tecnológica Federal do Paraná},
            city={Toledo},
            state={Paraná},
            country={Brasil}}

\affiliation[UNO]{organization={Universidade da Região de Chapecó},
            city={Chapecó},
            state={Santa Catarina},
            country={Brasil}}

\affiliation[IFPR]{organization={Instituto Federal do Paraná},
            city={Barracão},
            state={Paraná},
            country={Brasil}}

\affiliation[CEMHTI]{organization={Laboratoire CEMHTI, UPR 3079-CNRS},
            city={Orléans},
            country={France}}

\begin{abstract}
The glass industry has limited options to mitigate its environmental footprint, and the demand for cover glass to produce photovoltaic panels continues to increase. Currently, the majority of this special type of glass is not being recycled, therefore,  this work proposes to reuse it as raw material to obtain oxyfluoride glasses. The incorporation of CaF$_2$ and the increasing Na$_2$CO$_3$ content resulted in a melting temperature of about 1200$^\circ$C, significantly lower than in soda-lime glasses, contributing to the environmental benefits of reusing end-of-life cover glass. The obtained samples show high transparency and thermal stability, allowing the cover glass to make up to 80\% of its weight. XRF analysis was employed to determine the elemental composition of the samples, while XRD and Raman indicated that by adding CaF$_2$, the glass network was depolymerized. In situ XRD as a function of temperature showed the formation of a few crystalline phases in these oxyfluoride samples, evidencing their potential to be explored as a matrix to obtain different glass-ceramics. The combination of the glass properties indicates that this method and the resulting material can contribute to reducing the environmental impact of the glass industry. Furthermore, new glass or glass-ceramic materials can be obtained at a reduced temperature compared to soda-lime glass, while cover glass, being the primary raw material, could reduce the need to extract minerals from nature. 
\end{abstract}

\begin{graphicalabstract}
\end{graphicalabstract}

\begin{highlights}
\item Glasses containing up to 80\% cover glass content were obtained
\item Analyses have shown that introducing CaF$_2$ depolymerizes the material
\item These oxyfluoride glasses can be crystallized in several different crystalline phases
\item The samples exhibit transparency and thermal stability suitable for optical applications
\end{highlights}


\end{frontmatter}



\section{Introduction}
The world consumes over 130 million tons of glass annually, or about 16~kg per capita. Container and flat glass account for about 80\% of global glass production and are estimated to release more than 60 million tons of carbon dioxide into the atmosphere, primarily due to intensive heating of raw materials~\cite{furszyfer2022,zier2021}. Soda-lime is the most common glass, and the industry has limited options to mitigate its environmental footprint. To produce 1~kg of soda-lime glass, the carbonates present in the raw materials will emit around 0.15~kg of CO$_2$, while the huge amount of energy required to heat and melt ($\sim$1500$^\circ$C) the glass results in another 0.45~kg, primarily due to the combustion of fossil fuels~\cite{furszyfer2022,Butler2011}. Although clean electricity and hydrogen can be employed, these alternatives have yet to be demonstrated on an industrial scale~\cite{zier2021,caudle2023}.

Though glass can be recycled indefinitely, doing so is often not feasible or attractive due to technical and economic reasons~\cite{Westbroek2021,bristogianni2023}. As discussed in our previous work~\cite{Belancon2023}, highly transparent soda-lime is used as cover glasses (CG) in solar energy applications, and these high-quality materials are facing the risk of being dumped in landfills once their demand continues to rise, while the options to reuse them are pretty limited. Additionally, glass production requires vast amounts of raw materials and mining is one of the most energy-intensive industries~\cite{LI2024101597}, resulting in a severe environmental impact. These factors highlight the importance of recycling and reusing CG, turning end-of-life products into raw materials to keep them on the market, in a context of circular economy~\cite{Gaustad2018,Sica2018,Zink2019}.

Glasses other than soda-lime have been proposed as alternatives for solar energy applications, which could include new features such as spectral conversion to enhance silicon solar panel technology~\cite{Belancon2023}. However, among dozens of proposed glass systems, silicates are the only ones based on abundant, affordable, and non-toxic minerals. The resulting glasses often exhibit high transparency and adequate chemical and mechanical resistance for solar energy applications. Reviewing the literature recently~\cite{Belancon2023,FerrariMuniz2025}, we concluded that modified soda-lime silicates are the most promising system that could meet the requirements needed for mass-scale industrial production, while providing new or enhanced properties that could reduce industry energy consumption, expand solar power production, and contribute significantly to sustainability goals.

An interesting silicate system, which we have already investigated, is achieved when CaF$_2$ is added into silica melts~\cite{Muniz2021,Muniz2021a}. While modifiers such as CaO and Na$_2$O are well-known for reducing the melting temperature of silicates, the F provided by CaF$_2$ often replaces O, affecting the silica network~\cite{Zheng2025}. Several studies have shown that CaF$_2$ incorporation reduces the melting temperature and viscosity of silicates~\cite{Zheng2025}, while it can improve the microstructure~\cite{riaz2017} and promote or inhibit the formation of crystalline phases~\cite{Xu2025,Laonamsai2025}, including nanocrystals~\cite{Pawlik2021}. 

To address all the questions mentioned above, this work proposes using CGs from end-of-life solar panels as raw materials to produce oxyfluoride sodium calcium silicate glasses, achieved by incorporating CaF$_2$. Our methodology is simple; the obtained samples have high-optical quality and allow a recycled fraction of raw materials as high as 80\% of the final weight of the glass samples. Using CG as a raw material may contribute to the sustainability of the glass industry and positively impact the environment by reducing energy demand and the extraction of new raw materials from nature, ultimately reducing overall emissions and environmental footprint. Additionally, as we will show, these CG-derived materials may provide a platform to research and development of new glasses and glass-ceramics for several applications.

\section{Materials and methods}

The weight of the compounds mixed to produce each sample is shown in Table~\ref{tab:samples}, where the sample names refer to the mol\% of CaF$_2$. 
The CG, sourced from photovoltaic panels previously studied~\cite{Belancon2022a}, was fragmented into particles approximately 5–10~mm in length. The reagents added to the mix were CaF$_2$ and Na$_2$CO$_3$, both from Sigma-Aldrich with 99.0\% purity. To keep the ratio SiO$_2$/Na$_2$O approximately constant through all the samples, Na$_2$CO$_3$ concentration was adjusted accordingly.
\begin{table}[ht]
\caption{Mass of each sample prepared in this work}\label{tab:samples}
\centering
\begin{tabular}{c|c|c|c|c}\hline
  Sample & Cover-glass (g) & CaF$_2$ (g)& Na$_2$CO$_3$ (g)&Total (g)\\\hline \hline
  CgCAF05 & 13.00 & 1.00 & 1.94 &15.94\\
  CgCAF12 & 12.25 & 2.45 & 1.83 &16.53\\
  CgCAF15 & 11.80 & 3.00 & 1.76 &16.56\\
  CgCAF18 & 11.30 & 3.50 & 1.69 &16.49\\
  CgCAF20 & 10.90 & 4.00 & 1.63 &16.53\\\hline
\end{tabular}
\end{table}

\begin{figure}[ht]
\centering
\includegraphics[scale=0.30]{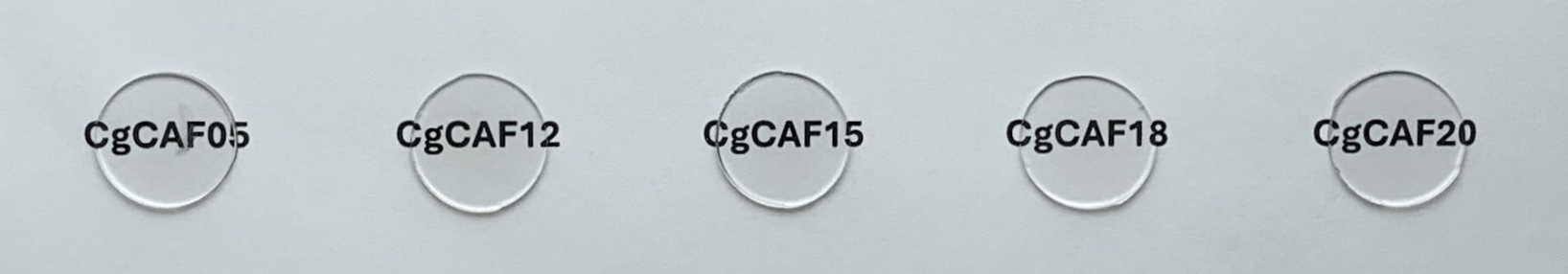}
\caption{Samples investigated in this work, as prepared for analyses. The diameter of the discs is about 1.0 cm.}
\label{fig:samples}
\end{figure}

Reagents were mixed with CG in a platinum crucible and heated in a furnace, using a 10 $^\circ$C/min heating rate, with holding steps of 40 minutes at 200$^\circ$C, 60 minutes at 500$^\circ$C, 90 minutes at 800$^\circ$C, and 120 minutes at 1200$^\circ$C. This thermal profile was chosen to ensure a homogeneous fusion of the mixture and to avoid foaming or spillage during melting. After this stage, the molten glass was poured into a stainless steel mold and rapidly transferred to a second furnace maintained at 480$^\circ$C, where it was kept for approximately 12 hours to relieve internal stresses. Upon cooling to room temperature, the samples were removed from the mold. Figure~\ref{fig:samples} shows a photo of the obtained samples, which were cut, polished, or milled for the analyses. 

The analytical techniques employed in this study encompass compositional, thermal, structural, and optical analyses. X-ray fluorescence (XRF) was used to determine the chemical composition of the samples. The analysis was performed on fused beads at approximately 1200$^\circ$C (Panalytical – Axios). The beads were prepared from powdered material, previously ground to pass through a 325-mesh sieve, using lithium tetraborate as the flux. Additionally, loss-on-ignition analysis was carried out to account for volatile content up to 1000$^\circ$C. Differential thermal analysis (DTA) was performed in the temperature range of 50$^\circ$C to 900$^\circ$C, using a heating rate of 10$^\circ$C/min under a synthetic air atmosphere with a flow rate of 50~mL/min (TA Instruments – SDT Q600). The density of the samples was obtained using Archimedes’ principle with distilled water at room temperature. Mass measurements were performed with an analytical balance with a precision of $\pm$ 0.1~mg (Kern ABT - 120-5DM).

Structural characterization was performed by X-ray powder diffraction (XRPD) on powdered samples. The analysis used CuK$\alpha$ radiation ($\lambda$= 1.5418~\AA) at 40~kV and 15~mA, with a scanning rate of 5$^\circ$/min, step size of 0.02$^\circ$, and a \(2\theta\) range of 5 to 80$^\circ$ (Rigaku - Miniflex 600). Also, in situ XRPD analysis was performed on powdered material to evaluate the temperature-dependent structural evolution. The measurements were conducted under air in the Bragg–Brentano $\theta$–$\theta$ geometry, using Ni-filtered CuK$\alpha_1$ radiation ($\lambda$ = 1.5418\AA), over a 2$\theta$ range of 15 to 60$^\circ$, with a step size of 0.016$^\circ$ and a counting time of 0.5 s/step. A temperature step of 25$^\circ$C was applied from 500 to 950$^\circ$C. Data were collected on a Bruker-AXS D8 Advance diffractometer (Karlsruhe, Germany) equipped with an Anton Paar oven chamber (model HTK 1200N, Graz, Austria), which allows measurements up to 1473 K. The powdered samples were placed in a platinum-lined corundum sample holder, and the diffracted intensities were recorded using a LynxEye detector.

Optical characterization involved Raman spectroscopy, UV-Vis-NIR and spectroscopic ellipsometry analysis. Raman spectra were collected on a Bruker-Senterra Raman microscope using a 532~nm laser at 20~mW, in the spectral range of 50 to 1542~cm$^{-1}$. A $\times$20 objective lens was used for laser focusing, and the signal was integrated for 2 seconds. UV-Vis-NIR transmittance was measured in the spectral range of 190 to 3600~nm (Shimadzu - UV-3600i Plus), and the refractive index of the samples was determined by
spectroscopic ellipsometry at an incidence angle of 55$^\circ$ (HORIBA - UVISEL Plus).

\section{Results and discussion}

Although the CG is expected to be low iron soda lime glass, and the reagents added are quite pure, the samples were submitted to XRF analysis to confirm their composition. These results are shown in Table~\ref{tab:frx}.
\begin{table}[ht]
\caption{Composition obtained by XRF expressed as a percentage (mol\%). *CaF$_2$ concentration was estimated assuming that Ca concentration obtained by XRF is the sum of CaO (from CG) plus CaF$_2$ (added).}\label{tab:frx}
\centering
\begin{tabular}{c|c|c|c|c|c|c}\hline
  Compound & CG& CgCAF05 & CgCAF12 & CgCAF15 &CgCAF18&CgCAF20\\\hline \hline
  SiO$_2$ &72.62 &62.90&58.22 &56.62 &54.18 &52.51\\
  CaO&9.35 &7.63 &7.59 &7.35 &7.11 &6.88\\
  Na$_2$O&12.48 &17.62 &16.12 &15.28 &15.19 &15.15\\
  MgO&4.46 &3.81 &3.55 &3.44 &3.29 &3.19\\
  Al$_2$O$_3$&0.48 &0.37 &0.36 &0.33 &0.31 &0.32\\
  K$_2$O&0.32 &0.26 &0.26 &0.22 &0.24 &0.24\\
  CaF$_2$*&- &5.20 &12.40 &15.25 &17.98 &20.61\\\hline
\end{tabular}
\label{tab:xrf}
\end{table}

Assuming the CG composition obtained from the XRF analysis and the weights shown in Table~\ref{tab:samples}, we can calculate the stoichiometric relations for CgCAF samples, which are very similar to those obtained experimentally by XRF. In this way, though fluorine is well known to be volatile, these results corroborate the assumption that no significant mass was lost during the melting process. As previously stated, we also aimed to keep the ratio SiO$_2$/Na$_2$O constant, and the XRF analysis confirms this it lies in the range of $\sim$3.5 - 3.7 for the samples.

\begin{figure}[ht]
\centering
\includegraphics[scale=1.4]{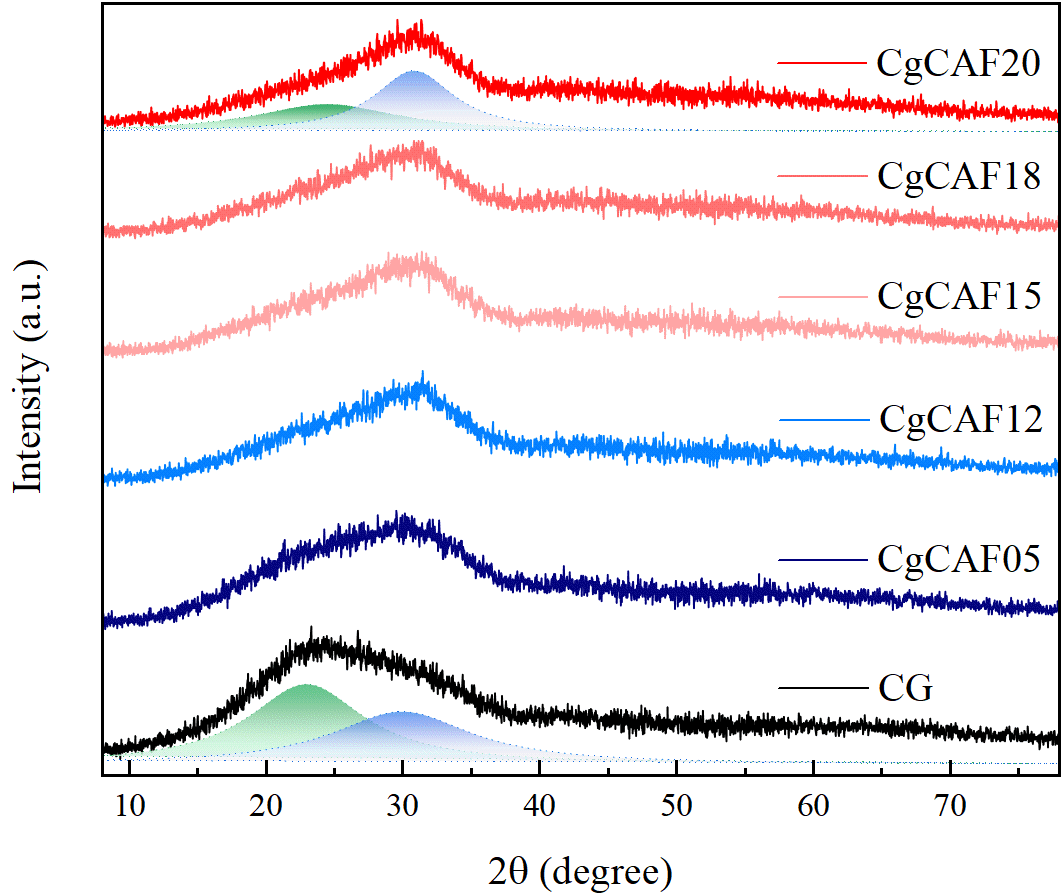}
\caption{X-ray powder diffraction patterns.}
\label{fig:drx}
\end{figure}

Figure~\ref{fig:drx} presents the XRPD patterns for all glass samples. The absence of sharp peaks in all patterns confirms the amorphous nature of the samples, consistent with a glassy structural framework. These results are in agreement with previously reported diffractograms for glasses and glass-ceramics of similar compositions~\cite{Muniz2021,almeida2008,Russel2005}.
Notably, the diffraction halos exhibit two overlapping broad bands, which can be more clearly distinguished by fitting the patterns with two Lorentzian curves, as shown in~\ref{fig:drx}, represented by the filled green and blue curves corresponding to bands centered at aprproximately 22$^\circ$ and 30$^\circ$, respectivelly. The area ratio between the green and blue bands decreases from 1.16 for the CG sample to 0.69 to the CgCAF20 sample.
In the CG sample, the most intense band is centered around 22$^\circ$, whereas in samples with increasing CaF$_2$ content, the band at 30$^\circ$ becomes more prominent. This behavior is likely associated with structural rearrangements driven by the incorporation of fluoride ions. As reported by Iwamoto and Makino~\cite{Iwamoto1981}, fluoride ions in calcium fluorosilicate systems can engage in Si-F or Ca-F bonding. At lower CaF$_2$ concentrations, the formation of Si-F bonds is favored, whereas at higher concentrations, Ca-F interactions become more prominent. This transition is accompanied by a change in the fluoride coordination environment, from four-fold coordination, characteristic of a CaF$_2$-type quasi-lattice, to six-fold coordination typical of a NaCl-type quasi-lattice. Notably, the NaCl-type structure possesses smaller average lattice parameters than the CaF$_2$-type configuration. According to Bragg’s law, a reduction in the lattice spacing results in a shift of diffraction features towards higher 2$\theta$ angles. Nonetheless, confirmation of this hypothesis requires more comprehensive structural analyses, including advanced modeling approaches. 

\begin{figure}[ht]
\centering
\includegraphics[scale=1.4]{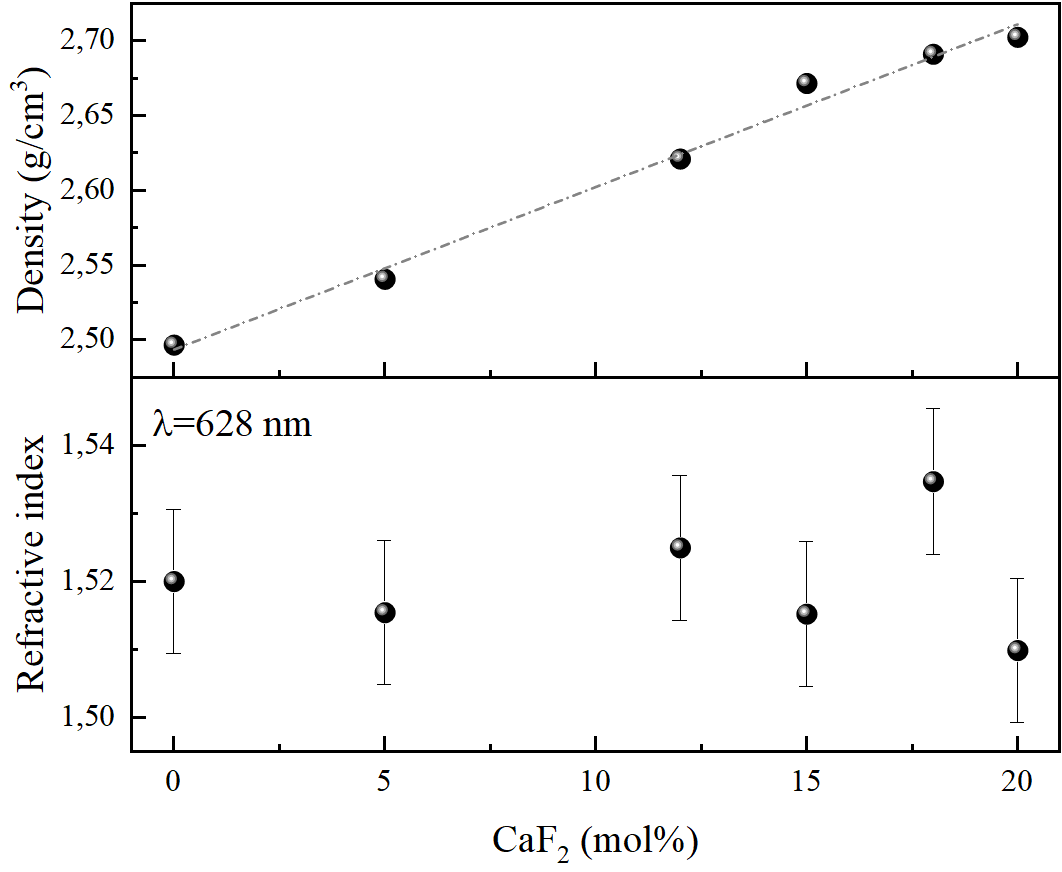}
\caption{Mass density and refractive index at 628~nm of the samples as a function of CaF$_2$ concentration. Error bars are negligible for the density in this range.}
\label{fig:den}
\end{figure}

As demonstrated in Table~\ref{tab:frx}, CgCAF samples have a significantly lower proportion of SiO$_2$ compared to the CG as CaF$_2$ concentration increases. Figure~\ref{fig:den}, presents the mass density and refractive index at 628~nm as a function of CaF$_2$ concentration.
The mass density of the samples increases linearly with the addition of CaF$_2$, with an experimental uncertainty of $\pm$0.001~g/cm$^3$, which renders the associated error bars negligible in the graphical representation. The linear increase in density observed with increasing CaF$_2$ content in the multicomponent soda-lime glass matrix arises from the combined effects of compositional modifications and structural rearrangements. Specifically, the gradual replacement of the lighter network-forming, SiO$_2$ by CaF$_2$, along with concurrent variations in the concentrations of other constituents such as Na$_2$O, CaO, and MgO, introduces heavier cations (e.g., Ca$^{2+}$, Na$^{+}$) and larger anions (F$^{-}$) into the glass structure. These changes lead to an increase in the average molar mass and promote a partial depolymerization of the silicate network, thereby enhancing atomic packing efficiency. The resulting structural compaction is corroborated by the shift of the amorphous halo to higher 2$\theta$ values in the XRPD patterns, indicating a reduction in the mean interatomic distances. At the same time, the refractive index remained constant ($n~\sim 1.52$), within the error margins. Muniz et al.~\cite{Muniz2021} have investigated sodium calcium silicate glasses containing amounts of CaF$_2$ similar to those in the present work, showing about the same mass density and a smaller refractive index (1.47). 

\begin{figure}[ht]
\centering
\includegraphics[scale=1.4]{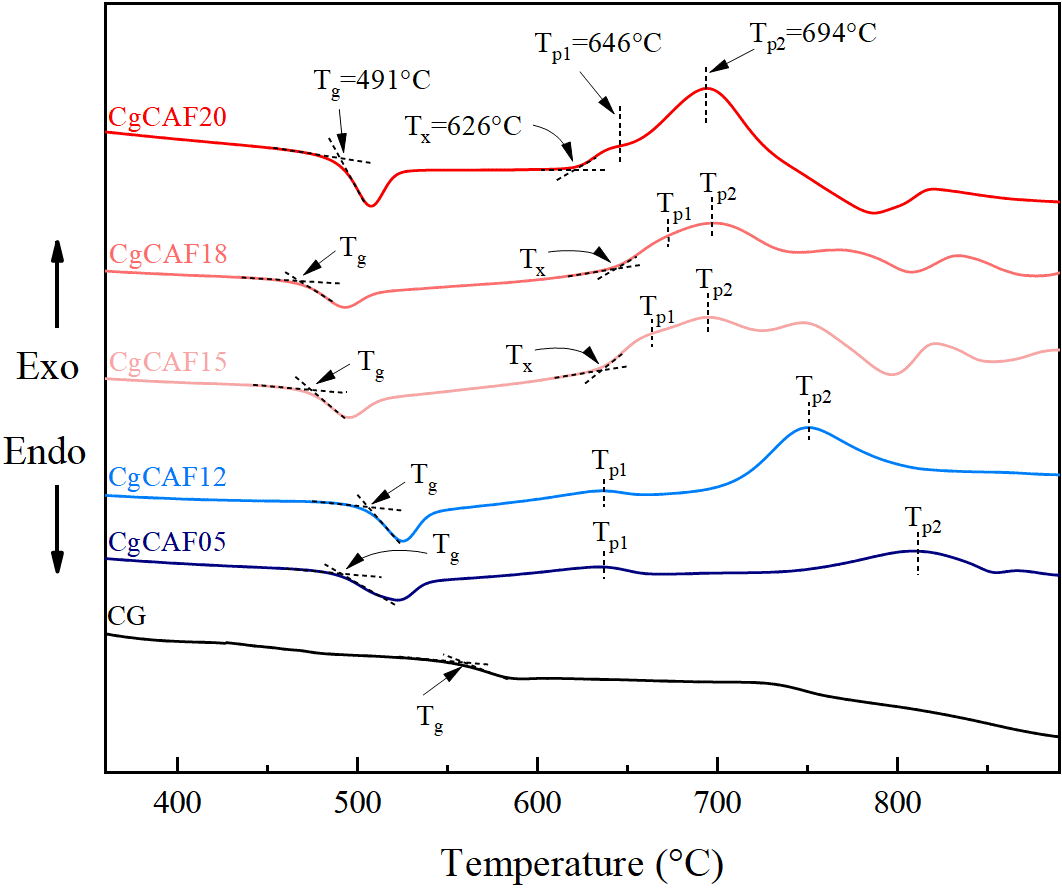}
\caption{Differential thermal analysis of CG and CgCAF samples.}
\label{fig:dta}
\end{figure}

Figure~\ref{fig:dta} presents the thermograms of all samples, including CG, where the key events T$_g$ (glass transition), T$_x$ (onset of crystallization), and T$_p$ (crystallization peak) are also indicated. A summary of the characteristic temperatures identified for all samples is shown in Table~\ref{tab:tg}. However, in these experiments, T$_x$ could not be identified for the CG, CgCAF05, and CgCAF12 samples.
As can be seen, in all CgCAF samples, it is possible to identify two crystallization peaks, a result widely observed in the literature on silicates containing CaF$_2$~\cite{Muniz2021,riaz2017}, as well as a significant reduction in the glass transition temperature ($\sim$50-90$^\circ$C).
\begin{table}[ht]
\centering
\begin{tabular}{c|c|c|c|c|c}
\hline
\textbf{Sample} & \textbf{T$_g$ ($^\circ$C)} & \textbf{T$_x$ ($^\circ$C)} & \textbf{T$_p(1)$ ($^\circ$C)} & \textbf{T$_p(2)$ ($^\circ$C)} & \textbf{T$_x$-T$_g$ ($^\circ$C)}\\ \hline \hline
CG & 560 & - & - & - & -\\ \hline
CgCAF05 & 491 & - & 636 & 812 & -\\ \hline
CgCAF12 & 507 & - & 637 & 751 & -\\ \hline
CgCAF15 & 475 & 636 & 664 & 695 & 161\\ \hline
CgCAF18 & 469 & 644 & 673 & 697 & 175\\ \hline
CgCAF20 & 491 & 626 & 646 & 694 & 135\\ \hline
\end{tabular}
\caption{Thermal parameters of glass samples.}
\label{tab:tg}
\end{table}
It is essential to note the satisfactory glass stability (T$_x$-T$_g$) observed in these samples, suggesting a suitable working temperature that may enable these materials to be produced in different shapes and sizes. Although T$_x$ could not be identified.

\begin{figure}[ht]
\centering
\includegraphics[scale=0.4]{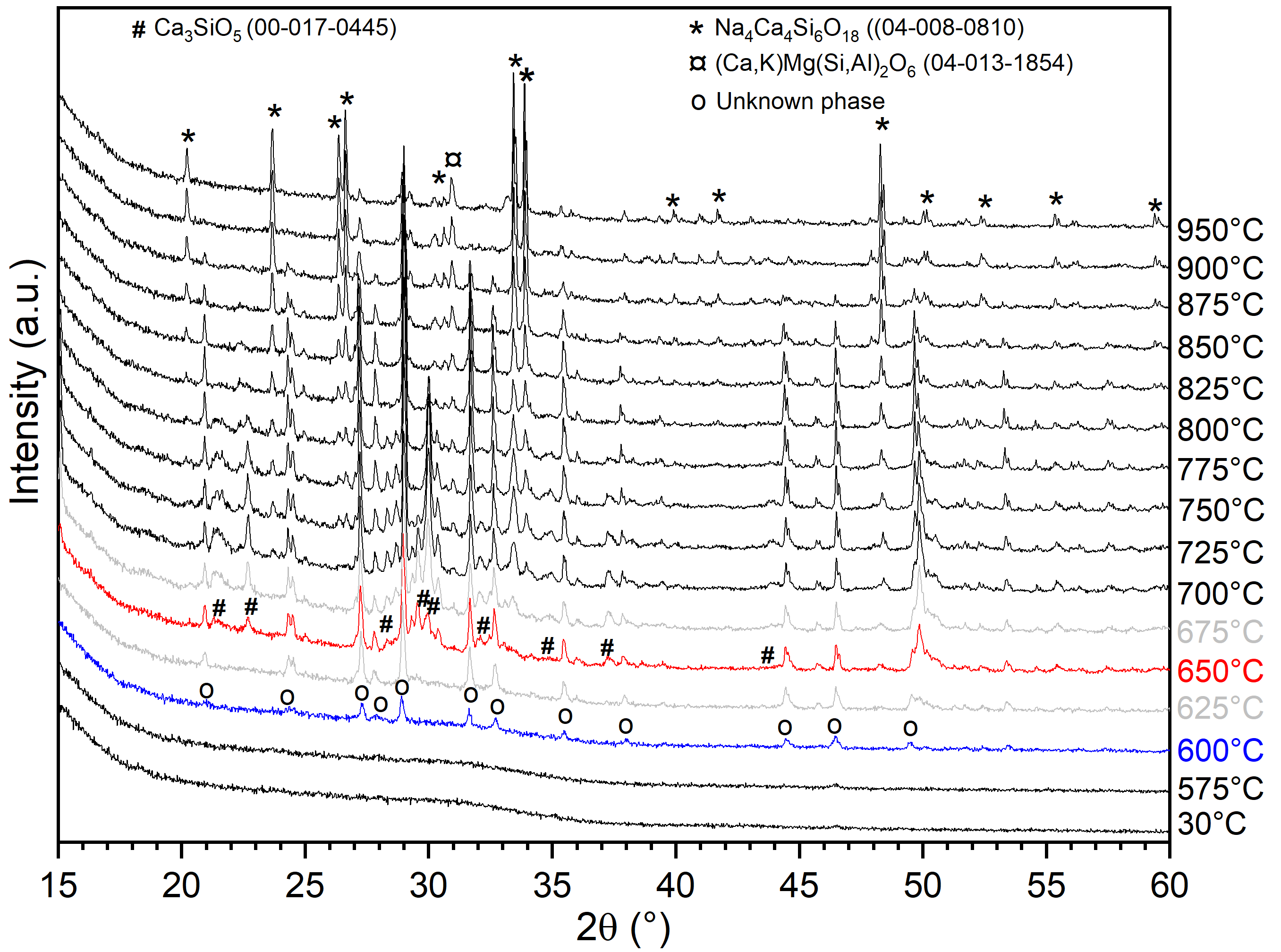}
\caption{In situ XRPD performed on the CgCAF12 samples.}
\label{fig:insitu}
\end{figure}

An in situ high-temperature XRPD study was performed on the CgCAF12 sample to investigate its crystallization characteristics under thermal treatment. The corresponding diffraction patterns are shown in Figure~\ref{fig:insitu}.

Consistent with DTA measurements, a crystalline phase appears at 600$^\circ$C. Nevertheless, we cannot identify this phase using the ICDD PDF5 database. Additional work will be necessary to elucidate this point.
The proportion of this unknown phase increases until 800$^\circ$C and after decreases but is present in small amounts at 950$^\circ$C. At 650$^\circ$C we observe the crystallization of C3S (Ca$_3$SiO$_5$ JCPDS file 00-017-0445), which disappears at 800$^\circ$C. The Combeite phase (Na$_4$Ca$_4$Si$_6$O$_{18}$ JCPDS file 04-008-0810) crystallizes around 700$^\circ$C, and its proportion increases until 950$^\circ$C to be finally the main phase at high temperature. After this experiment, the PXRD diagram collected at room temperature shows the presence of Combeite and a minimal amount of Akermanite (JCPDS file 01-079-2425). Some of these phases have been linked to intriguing features, such as antimicrobial activity~\cite{Caland2024}. This indicates that CG-derived materials may also result in glass-ceramics with unique properties, which we will explore in future work.

\begin{figure}[ht]
\centering
\includegraphics[scale=1.4]{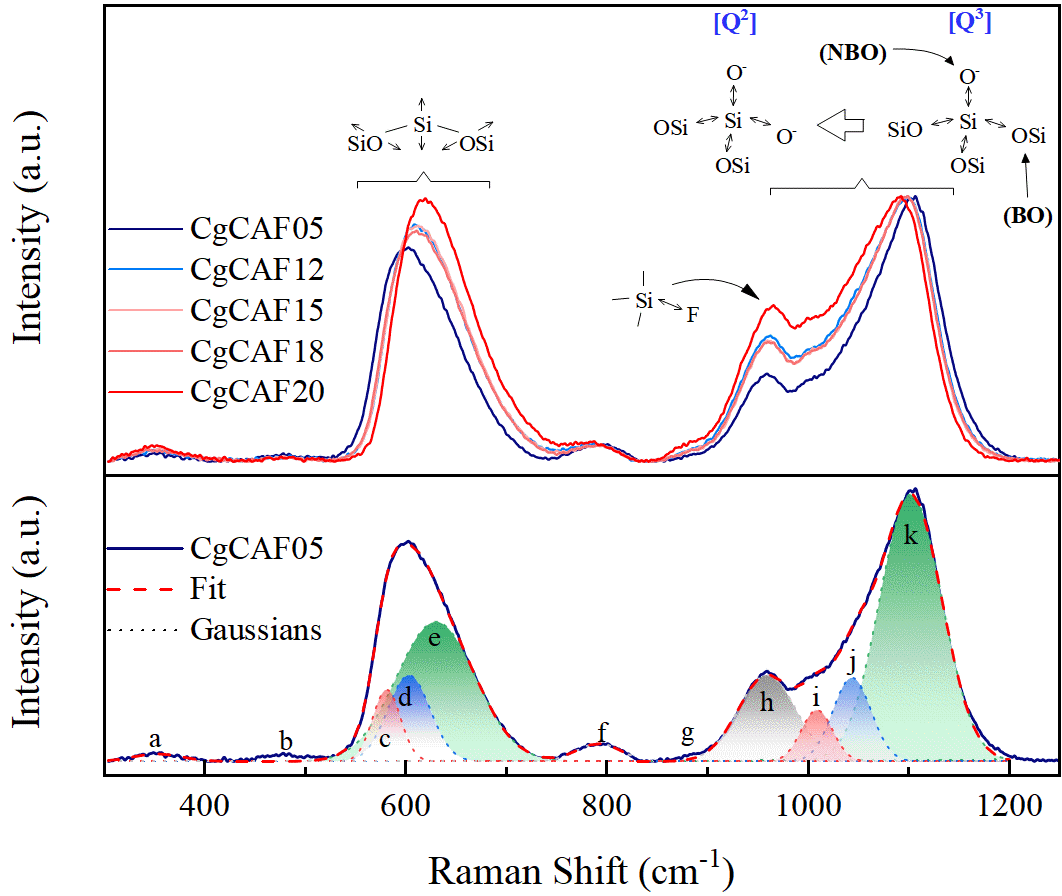}
\caption{Raman spectra of CgCAF samples.}
\label{fig:raman}
\end{figure}

Figure~\ref{fig:raman} shows the Raman spectra recorded for all samples, along with deconvolution using a multipeak Gaussian fit applied to the CgCAF05 sample (including the individual Gaussian components) as a representative example. Each Raman spectrum was fitted using 11 Gaussian components labeled (a to k) in the figure. This number was chosen based on the stability of the fitting process, as additional components did not significantly improve the quality of the fit. A weak band (g), located at approximately 870~cm$^{-1}$, could not be fitted for the CgCAF05 sample. The Raman scattering of the studied glasses is dominated by bands in the 550–750~cm$^{-1}$ range (c, d and e) and in the 900–1200~cm$^{-1}$ range (h, i, j and k). Raman spectroscopy has proven to be a powerful technique for the structural investigation of silicate glasses~\cite{Vidal2016, Gao2016, Tian2022}, although multiple overlapping bands pose a significant challenge for complete spectral interpretation.

The bands within the 900–1200~cm$^{-1}$ range (h, i, j and k) are attributed to symmetric stretching vibrations of Si–O bonds in various [SiO$_4$] tetrahedral units, commonly referred to as Q$^n$ bands (where $n$ = 0 to 4), with $n$ denoting the number of bridging oxygens per tetrahedron~\cite{Yadav2015, Tian2022}. It is well established that the band near 1100~cm$^{-1}$, associated with the Si-O vibration involving bridging oxygens, shifts to lower frequencies - or new bands emerge in lower frequency regions - due to the disruption of Si–O–Si linkages~\cite{Tsunawaki1981}. Furthermore, the primary determinant of Raman band frequency within this range is the force constant of the Si–NBO (non-bridging oxygen) bond. In silicate glasses containing CaF$_2$, the substitution of oxygen atoms by fluorine distorts the silicon's local electronic environment due to the higher electronegativity of fluorine. This distortion weakens the remaining Si–O bonds within the tetrahedra, leading to lower bond force constants and, consequently, lower associated vibrational frequencies~\cite{Luth1988}. This effect explains the slight ``k" band shift with increasing CaF$_2$ concentration; this behavior was also observed by Muniz et al.~\cite{Muniz2020} for a similar glass composition.

On the other hand, the Si-F bond behaves similarly to the Si-O bond, and the stretching vibration of the Si-F bond in SiO$_3$F$^-$ tetrahedra typically appears near 950~cm$^{-1}$ in fluorine-containing silica glasses~\cite{Luth1988, Dumas1982}. Therefore, the formation of Si-F bonds may influence the frequencies of Si-O related vibrations and could account for the systematic increase in intensity observed for band ``h" ($\sim$960~cm$^{-1}$) as CaF$_2$ content increases.

The bands in the 600~cm$^{-1}$ region (c, d, and e) are assigned to symmetric vibrations of bridging oxygens (BO) in Si-O-Si linkages~\cite{Yadav2015, Muniz2020}. In conventional soda-lime glasses, bands near 500 and 560~cm$^{-1}$ are attributed to bending motions of Si-O-Si in highly polymerized tetrahedral structures (Q$^4$ and Q$^3$ species), while bands above 600~cm$^{-1}$ are typically associated with Si-O bending in depolymerized [SiO$_4$] tetrahedra (Q$^2$ species)~\cite{Yadav2015, Wang2011}. Although the Raman spectra of the intermediate samples (with CaF$_2$ contents ranging from 12.15 to 18~mol\%) exhibit negligible variation among themselves, a notable shift of approximately 20~cm$^{-1}$ is observed in the position of bands ``e'' and ``k'' when comparing the initial (CgCAF05) and final (CgCAF20) samples of the series. This spectral evolution indicates an increase in the fraction of NBO relative to total oxygens, suggesting a progressive depolymerization of the silica network with increasing CaF$_2$ content~\cite{Muniz2021, Novatski2008, OShaughnessy2020}. Such behavior implies significant structural modifications, including the possible formation of new Q$^n$ species or phases, or a substantial reorganization within the silicate network.

The observed low-intensity Raman bands can also be identified. Band ``a'', located near 350~cm$^{-1}$, is attributed to the vibrations of the network-modifying cations~\cite{Muniz2020}. Although weak, its integrated area increases linearly with CaF$_2$ concentration. Band ``f'', observed at 800~cm$^{-1}$ and invariant concerning CaF$_2$ content, is assigned to the Si-O stretching with dominant Si atom motion~\cite{Boizot2003}. Symmetric stretching vibrations of Ca-F bonds are expected to appear around 485~cm$^{-1}$ (band ``b'')~\cite{Luth1988}. However, due to the highly ionic character of the Ca-F bond, Raman bands associated with these vibrations exhibit low intensity, and barely to no variation is observed as a function of CaF$_2$ content in the glass compositions studied.

\begin{figure}[ht]
\centering
\includegraphics[scale=1.4]{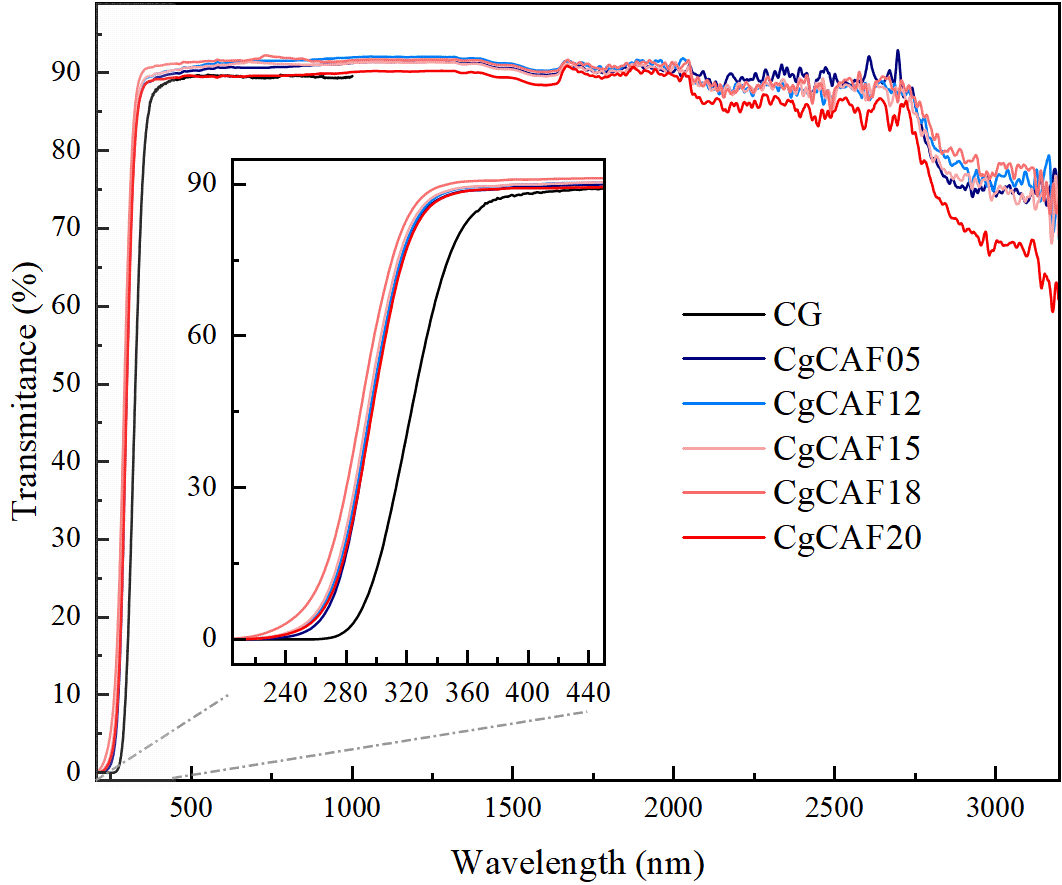}
\caption{UV-VIS spectra.}
\label{fig:uvvis}
\end{figure}

Finally, samples of about 0.6~mm were used to measure the light transmittance in the range 190-3500~nm. These spectra are shown in Figure~\ref{fig:uvvis}, demonstrating the overall high transmittance of the CgCAF samples. 
As shown in Figure~\ref{fig:den}, the refractive index of these samples is $n\cong1.5$, which translates to reflection losses of about 4\% at each air-glass interface. As can be seen, the transmittance remains near 90\% from UV to NIR, evidencing the high optical quality of the CgCAF samples in this range. As transmittance ($\sim$90\%) plus reflection ($\sim$8\%) accounts for about 98\% of the incoming light, it can be concluded that the absorption and scattering losses are pretty low in these samples. The inset in Figure~\ref{fig:uvvis} also highlights the wider transmittance window in the UV range of CgCAF samples compared to the original soda-lime glass, which may be associated with changes in the glass structure and could point to potential applications in UV-related technologies.

\section{Conclusions}
In this work, we have demonstrated the utilization of end-of-life cover glass from commercial solar panels to produce oxyfluoride glasses by incorporating CaF$_2$. XRF analyses have proven that the final samples are free of contaminants, such as iron, that could introduce color to the material, reducing its value and limiting its potential applications. This confirms that, due to the high purity and transparency of the soda-lime cover glass, we could use it to make up to 80\% of the total weight of CgCAF glasses. In contrast, conventional flat glass production allows only a small proportion of the material to be recycled. Furthermore, the melting temperature of the samples was about 1200$^\circ$C, significantly lower than the melting temperature of soda-lime glass, which favors a reduced energy consumption and carbon emissions to produce CgCAF samples. Thermal analysis showed a 50-90$^\circ$C reduction in T$_g$ temperature relative to the soda-lime, the presence of at least two crystallization processes and a pretty good stability of the glass phase, demonstrated by (T$_x$-T$_g$) values in the range 100-180$^\circ$C. XRPD confirmed the glassy nature of the samples and indicated a depolymerization process as the CaF$_2$ concentration increases, which is corroborated by mass density and Raman data. In situ XRPD measurements in the CgCAF12 sample demonstrated the formation of combeite, diopside, and Ca$_3$SiO$_5$ triclinic phases, besides some unidentified peaks that require further studies. Finally, the UV-VIS-NIR transmittance demonstrated the good optical quality of all samples and the improved UV transmittance window compared to the original cover glass employed in this work. The ensemble of results shows that this family of materials can be explored as glass or be crystallized to produce glass-ceramics that may be tailored for different applications. Further studies should focus on understanding the crystallization process of these materials and incorporating optically active ions. If scaled for practical applications, these oxyfluoride glasses could provide a viable route for the reintegration of end-of-life cover glass into the industrial value chain. Such an approach would not only enhance the economic value of this waste material but also promote circularity and sustainability within the glass and photovoltaic industries.

\section{Acknowledgement}

The authors thank CNPq (grants 409475/2021-1, 402473/2023-0 and 304060/2023-2) for the financial support.

\bibliographystyle{elsarticle-num} 
\bibliography{references}

\end{document}